\documentclass{aa}
\usepackage{graphicx}
\usepackage{txfonts}
\begin{document}
   \title{
   How young stellar populations affect the ages and metallicities of galaxies
   \thanks{The main results are only available in electronic form
   at the CDS via anonymous ftp to cdsarc.u-strasbg.fr (130.79.128.5)
   or via http://cdsweb.u-strasbg.fr/cgi-bin/qcat?J/A+A/}}


   \author{Zhongmu Li
          \inst{1, 2}
          \and
          Zhanwen Han\inst{1}
          }

   \offprints{Zhongmu Li}

   \institute{National Astronomical Observatories/Yunnan Observatory, the Chinese Academy of Sciences, Kunming, 650011,
   China\\
              \email{zhongmu.li@gmail.com}
         \and
            Graduate School of the Chinese Academy of Sciences\\
             }
   \titlerunning {Effects of YSPs on determination of stellar populations}
   \date{Received September 15, 1996; accepted March 16, 1997}


  \abstract
   {Tow important stellar-population parameters (age and metallicity) of the dominant stellar populations (DSPs) of galaxies are usually estimated by comparing the observed absorption line
   indices or colors to predictions of
   some simple stellar population models.
   However, some studies show that there is actually recent star formation in galaxies, including early type ones. This suggests that
   we may not be obtaining accurate the two stellar-population parameters for the DSPs of galaxies. This is obvious when we estimate the
   two parameters by colors, because the youngest
   populations dominate the light and make the fitted stellar populations younger and richer in metal.}
   {We plan to study how young populations (YSPs) in composite stellar
   populations (CSPs) affect the colors of star systems and to analyze how the
   stellar ages and metallicities derived from colors possibly deviate
   from those of the DSPs.}
   {The techniques of stellar population synthesis and Monte Carlo process are used in this analysis.}
   {It is found that the age and mass fraction of a YSP affect colors of a mixed star system significantly, but the former is stronger.
   In addition, our results show that the stellar ages and metallicities derived directly from a pair of colors
   are about 2.14 Gyr younger, while 0.0027 more metal rich on average than those of the DSPs of composite stellar systems.
   Some possible distributions of the differences between stellar-population parameters determined by colors and
   those of DSPs of CSPs are presented.
   The possible distributions of the differences between colors of CSPs and
   those of their DSPs are also shown. Stellar ages and metallicities measured by colors and line-strength indices are compared in the work,
   with a sample of 18 galaxies. Furthermore, the YSPs may affect the fundamental plane and Kormendy relation of early type galaxies.
   }
   {}


   \maketitle
%

\section{Introduction}

    To study the formation and evolution of galaxies, we usually need to estimate two stellar-population parameters (age and metallicity) of
    the dominant stellar populations (DSPs), which contribute most to the stellar mass of galaxies.
    The two parameters of DSPs of galaxies without obvious star formation, e.g., early type galaxies,
    are usually estimated by comparing the observed values with the predictions of simple stellar populations
    (SSPs)
    (see, e.g., Worthey \cite{worthey94}; Vazdekis \cite{vazdekis99}; Bruzual \& Charlot \cite {bruzual03} (hereafter
    BC03);
    Fioc \& Rocca Volmerange \cite{fioc97} (PEGASE code); Zhang et al. \cite{zhang05}).
    Most of these works are accomplished by measuring absorption line indices or spectra, e.g., Kuntschner (\cite {kuntschner00}),
    Vazdekis et al. (\cite {vazdekis97}), Gallazzi et al. (\cite{gallazzi05}), Li et al. (\cite{li06}), Zhou et al. (\cite{zhou92}),
    Kong \& Cheng (\cite{kong98}), and Kong et al. (\cite{kong03}).
    A few works have also
    tried to obtain similar results
    by measuring colors, e.g., Dorman et al. (\cite{dorman03}), Yi (\cite{yi03}), Wu et al. (\cite{wu05}), James et al. (\cite{james06}),
    Li et al. (\cite{li07}), Kaviraj et al. (\cite{kaviraj06}), and Lee et al. (\cite{lee06}),
    because colors are very useful for studying the stellar populations of distant galaxies.
    However, many studies show
    that there is recent star formation in galaxies of all
    types (Trager et al. \cite{trager00}; Yi et al. \cite{yi05}; Schawinski et al. \cite{schawinski06}),
    which suggests that we have not
    obtained accurate ages and metallicities of the DSPs of galaxies, via SSPs
    with instantaneous star burst (see, e.g.,
    Serra \& Trager \cite{serra07}; Li et al. \cite{li07}). This is true
    when one uses colors to estimate the two parameters. If
    there are two populations in galaxies, we will usually get younger and
    more metal-rich stellar populations than the DSPs of galaxies.
    Thus it is difficult to get reliable estimates for the ages and metallicities of
    DSPs of galaxies from colors before we know how much the two values derived from colors are possibly different from those of their
    DSPs. Unfortunately, there has been no detailed study of this yet.

    We intend to get some answers in this work. The main aims of the paper are to study how the
    parameters relating to the young population (YSP) of a composite stellar population (CSP) affect the colors of the
    system and to analyze the distributions of the deviations in stellar ages and metallicities
    determined by the colors of CSPs when comparing them to their DSPs. We also intend to
    study the possible distributions of the differences between colors of CSPs and their DSPs.

    The structure of the
    paper is as follows. In Sect. 2, we briefly introduce the construction of CSPs. In Sect. 3,
    we analyze how the parameters of the YSP of a CSP affect the colors of the system. In Sect. 4, we study
    the distributions of the differences between colors of CSPs and those of their DSPs. In Sect. 5,
    we give some possible distributions of differences between stellar ages and metallicities
    estimated via colors of CSPs and those of their DSPs. Then we try to study how
    we can correct for the deviations in
    the two stellar-population parameters, which result from the existence of YSPs in
    galaxies. The stellar ages and metallicities estimated by
    different methods (the color-method and absorption-line index method, a color
    pair method and multi-color method) are also compared in the
    section.
    Finally, we give our discussions and conclusions in Sect. 6.


\section{Construction of composite stellar populations }

    We construct CSPs using the SSPs of BC03, with a wide range in
    age (0.01 -- 20 Gyr) and metallicity (0.0001 -- 0.05).
    Each CSP is constructed of two SSPs: an old DSP and a YSP.
    There are five input parameters for each CSP: metallicity and age of the old DSP ($Z_{\rm 1}$ and $t_{\rm 1}$),
    metallicity and age of the YSP ($Z_{\rm 2}$ and $t_{\rm 2}$), and the mass fraction
    of the YSP in the total system ($F_{\rm 2}$). For the convenience of comparing our results
    with others, we mainly use [$Z$/H]$_{\rm 1}$=log($Z_{\rm 1}$/$Z_{\odot}$) and [$Z$/H]$_{\rm 2}$=log($Z_{\rm 2}$/$Z_{\odot}$)
    to represent the stellar metallicities of the DSPs and YSPs in this work. The range of $F_{\rm 2}$ is from 0.5\% to 50\%.
    Note that $t_{\rm 2} < t_{\rm 1}$, $Z_{\rm 2} \geq Z_{\rm 1}$.
    In this way, we construct 1\, 894\,200 CSPs and then calculate all their $UBVRIJHK$
    colors. The ranges of input parameters of the sample CSPs are listed in Table 1.

\begin{table}[]
\caption[]{Input parameters for modeling composite stellar
populations.} \label{Tab:1}
\begin{center}\begin{tabular}{ll}
\hline \hline\noalign{\smallskip}
parameter & values\\    
\hline\noalign{\smallskip}
[$Z/{\rm H}$]$_{\rm 1}$            & -2.30,-1.70, -0.70, -0.40, 0.0, 0.40\\
$t_{\rm 1}$ (Gyr)      & 0.05, 0.1, 0.51, 1.02, 1.61, 2, 2.5, 3, ..., 20\\
$[Z/{\rm H}]_{\rm 2}$           &  from $[Z/{\rm H}]_{\rm 1}$ to 0.40\\
$t_{\rm 2}$ (Gyr)      &  from 0.01 to $t_{\rm 1}$\\
$F_{\rm 2}$=M$_2$/(M$_1$+M$_2$) &  0.005, 0.01, 0.015, ..., 0.5\\
\noalign{\smallskip}\hline\hline
\end{tabular}\end{center}
\end{table}


\section{Effects of the YSP parameters on colors}

    A young population usually makes the whole star system bluer than the DSP of the system and
    results in incorrect determinations of stellar age and metallicity.
    However, we do not know how the parameters of the YSP affect the colors
    of the whole system.
\subsection{Effect of the age of the young population}

    We test the effect of the age of the YSP of a CSP, $t_{\rm 2}$, on the colors of the CSP via mixing
    YSPs with variable ages into the same DSP. Then we investigate how the
    differences between the colors
    of CSPs and those of the DSP change with the
    age of the YSP. We take the same metallicity for the YSP
    and DSP of a CSP here. The results for three colors that have the potential to disentangle the
    stellar age-metallicity degeneracy (Li et al. \cite{li07}),
    i.e., $(B-V)$, $(B-K)$, and $(I-H)$, are plotted in Figs. 1, 2,
    and 3, respectively. The differences in the colors, which are represented by $\Delta(B-V)$,
    $\Delta(B-K)$, and $\Delta(I-H)$, etc., are calculated by
    \begin{equation}
    \rm
        \Delta \it C = C_{\rm CSP} - C_{\rm DSP},
    \end{equation}
    where $\Delta \it C$ is the difference, while $C_{\rm CSP}$ and $C_{\rm DSP}$ are the colors
    of the CSP and DSP, respectively.
    We take a few CSPs with log($t_{\rm 1}/{\rm yr}$) of 10.1761 ($t_{\rm 1}$=15 Gyr) or 10.0 ($t_{\rm 1}$=10 Gyr)
    and metallicity ([$Z$/H]$_{\rm 1}$=[$Z$/H]$_{\rm 2}$) of 0.0 or 0.40 in this study.
    Four mass fractions (0.5\%, 1.0\%, 5.0\%, 10.0\%) are chosen for
    the YSPs in the tests.
    As we see, the results for different colors are similar: The differences between
    colors of a composite star system
    and those of the DSP of the system increase rapidly
    with decreasing age of the YSP when
    the age is less than about 5 Gyr.
    The results for some other colors and other CSPs, which are not shown in the paper,
    are found to be similar to those shown here.

   \begin{figure}
   \centering
   \includegraphics[angle=-90,width=88mm]{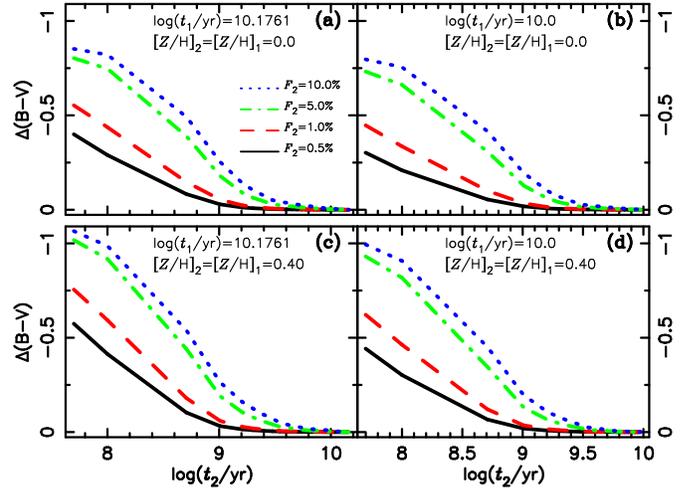}
      \caption{Differences between $(B-V)$ colors of a CSP and DSP versus YSP age ($t_{\rm 2}$). Solid, dashed, dot-dashed, and
      dotted lines correspond to YSP mass fractions ($F_{\rm 2}$) of 0.5\%, 1.0\%, 5.0\%, and
      10.0\%, respectively.  $\Delta(B-V)$ is calculated by Eq. (1).
              }
         \label{FigVibStab}
   \end{figure}
%

   \begin{figure}
   \centering
   \includegraphics[angle=-90,width=88mm]{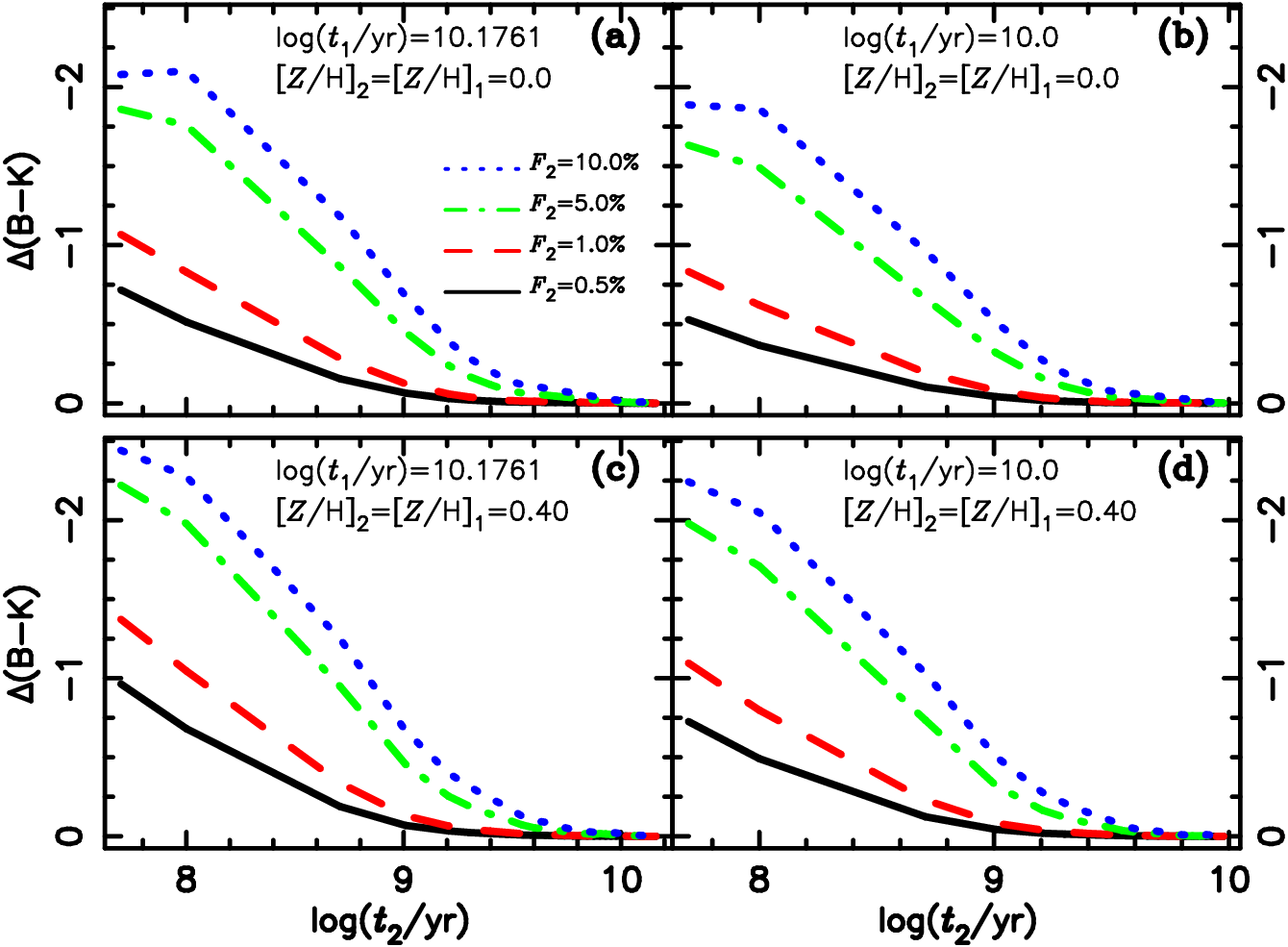}
      \caption{Similar to Fig. 1, but for $(B-K)$.
              }
         \label{FigVibStab}
   \end{figure}
%

   \begin{figure}
   \centering
   \includegraphics[angle=-90,width=88mm]{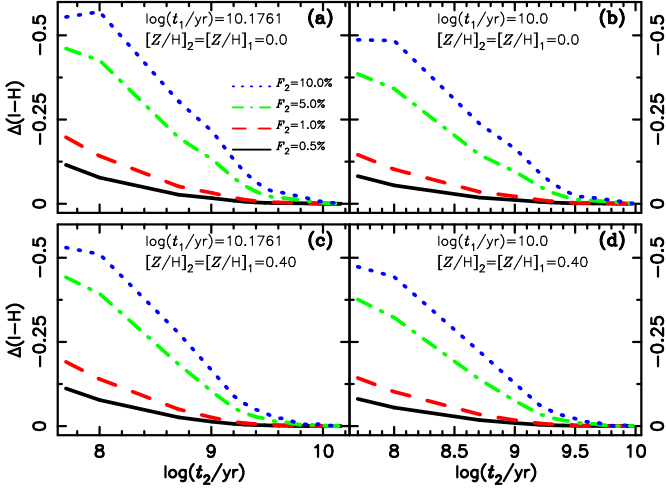}
      \caption{Similar to Fig. 1, but for $(I-H)$.
              }
         \label{FigVibStab}
   \end{figure}
%

\subsection{Effect of the mass fraction of the young population}

    The effect of the mass fraction of the YSP, $F_{\rm 2}$, on
    the colors of a CSP is studied by a method  similar to what was used to study the effect of the
    age of the young population. There $\Delta(B-V)$, $\Delta(B-K)$, and $\Delta(I-H)$ are shown to
    increase with $F_{\rm 2}$.
    We plot the results in Figs. 4, 5, and 6, respectively.
    We take 15 Gyr (log($t_{\rm 1}/{\rm yr}$)=10.1761) or 10 Gyr (log($t_{\rm 1}/{\rm yr}$)=10.0) for
    the ages of the DSPs and take 0.51 Gyr (log($t_{\rm 2}/{\rm yr}$)=8.7076) or 2 Gyr (log($t_{\rm 2}/{\rm yr}$)=9.3010)
    for the YSPs here.
    In each test, we adopt metallicities ([$Z$/H]$_{\rm 1}$=[$Z$/H]$_{\rm 2}$) of -1.70, -0.70, 0.0,
    and 0.40 for the CSPs, which are shown differently in the figures.
    The metallicities of the YSP and DSP of
    a CSP are assumed to be the same.
    We see that the results for three colors are similar:
    (1) Compared to those of DSPs, colors change quickly
    with increasing mass fraction of the young population, especially
    for mass fractions lower than 10\%.
    (2) The rate of change in the colors does
    not mainly depend on the difference between $t_{\rm 2}$ and $t_{\rm 1}$, but on the age of the YSP.

   \begin{figure}
   \centering
   \includegraphics[angle=-90,width=88mm]{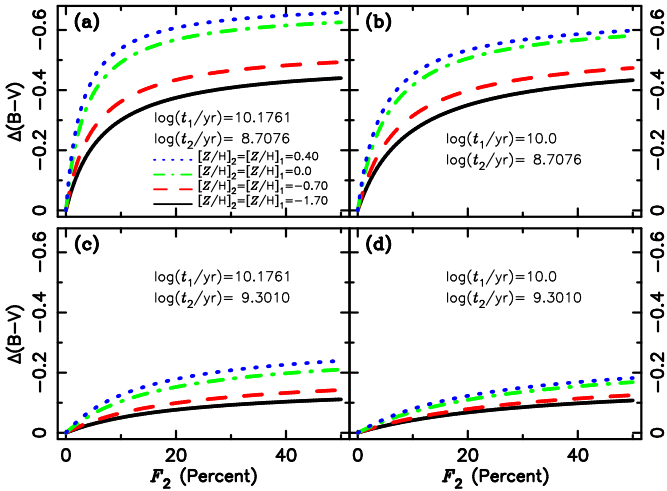}
      \caption{Difference between $(B-V)$ of a CSP and that of its
      DSP versus $F_{\rm 2}$. Solid, dashed, dashed-dotted, and
      dotted lines show CSPs with metallicities ([$Z$/H]$_{\rm 2}$=[$Z$/H]$_{\rm 1}$) -1.70, -0.70, 0.0,
    and 0.40, respectively. $\Delta(B-V)$ is calculated by equation (1).
              }
         \label{FigVibStab}
   \end{figure}
%

   \begin{figure}
   \centering
   \includegraphics[angle=-90,width=88mm]{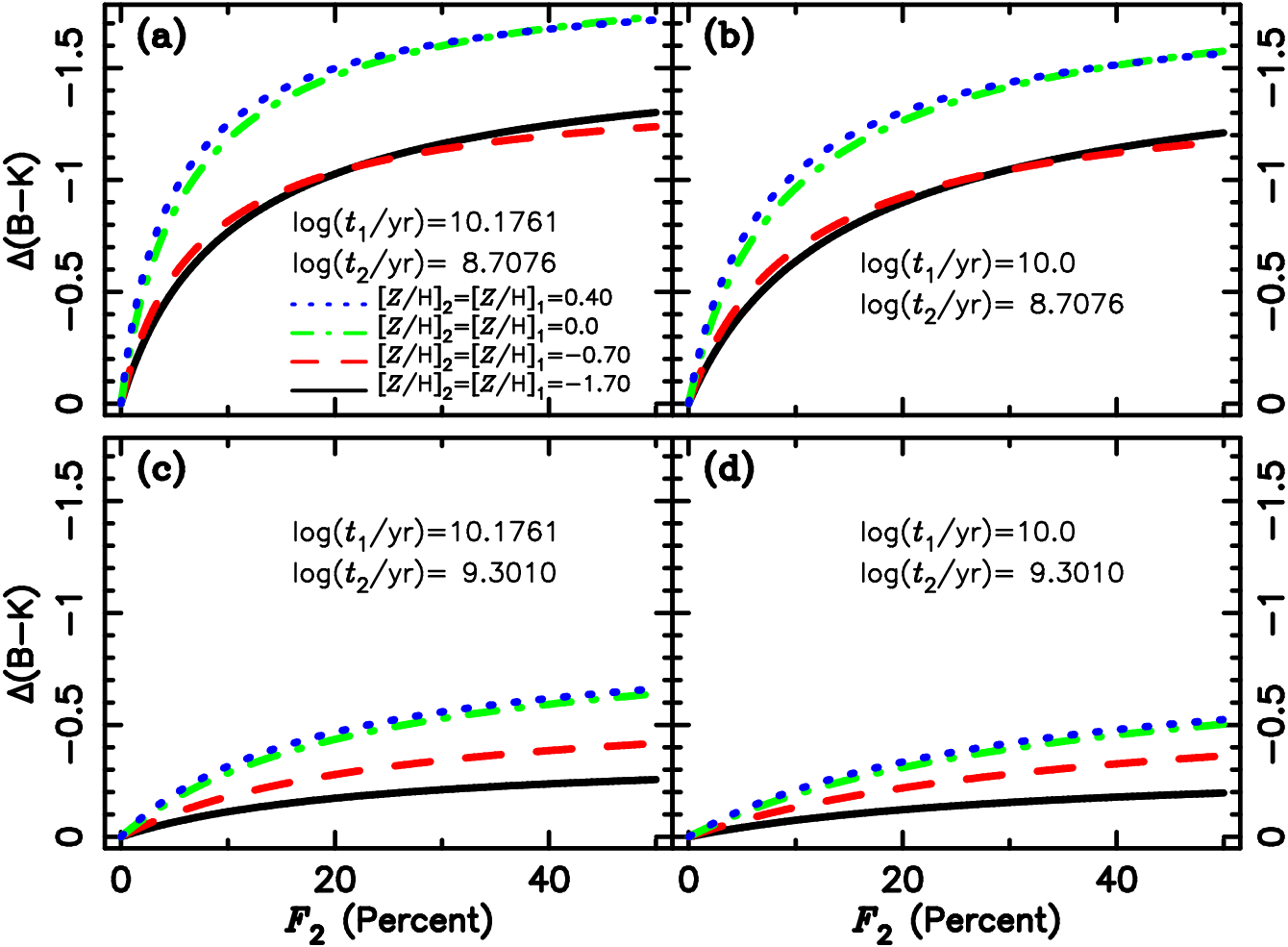}
      \caption{Similar to Fig. 4, but for $(B-K)$.
              }
         \label{FigVibStab}
   \end{figure}
%

   \begin{figure}
   \centering
   \includegraphics[angle=-90,width=88mm]{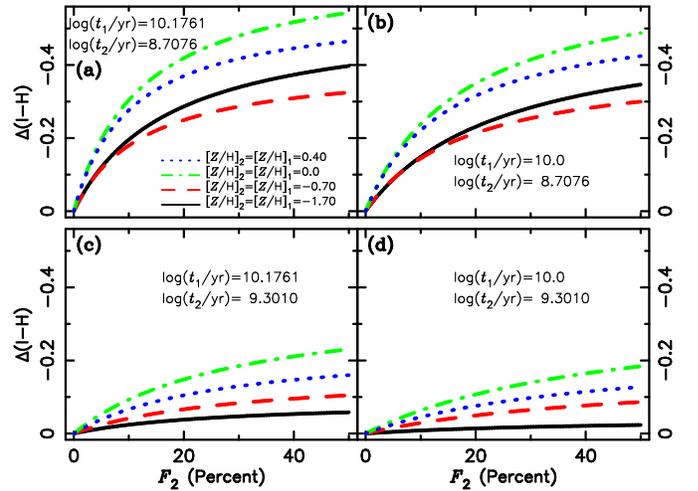}
      \caption{Similar to Fig. 4, but for $(I-H)$.
              }
         \label{FigVibStab}
   \end{figure}
%

\subsection{Effects of the metallicity of the young population}

    We also investigated the effects of the metallicity of the young population, i.e., [$Z/{\rm H}$]$_{\rm 2}$, on the colors of
    a CSP. We use a similar method
   to the one above. Some parameters, i.e., $t_{\rm 1}$, [$Z/{\rm H}$]$_{\rm 1}$, $t_{\rm 2}$, and $F_{\rm 2}$ are fixed
    for CSPs, while [$Z/{\rm H}$]$_{\rm 2}$ are variable in each test. Note that we assume
    [$Z/{\rm H}$]$_{\rm 2}$ $\geq$
    [$Z/{\rm H}$]$_{\rm 1}$. The results for $(B-V)$ are shown in Fig. 7.
    We see that [$Z/{\rm H}$]$_{\rm 2}$ or $Z_{\rm 2}$ does not affect the colors of star systems as
    clearly as $t_{\rm 2}$
    and $F_{\rm 2}$, but the effect is obvious for CSPs with metal-rich DSPs.
    The results for other colors are not shown here,
    since they are similar to those of $(B-V)$.

   \begin{figure}
   \centering
   \includegraphics[angle=-90,width=88mm]{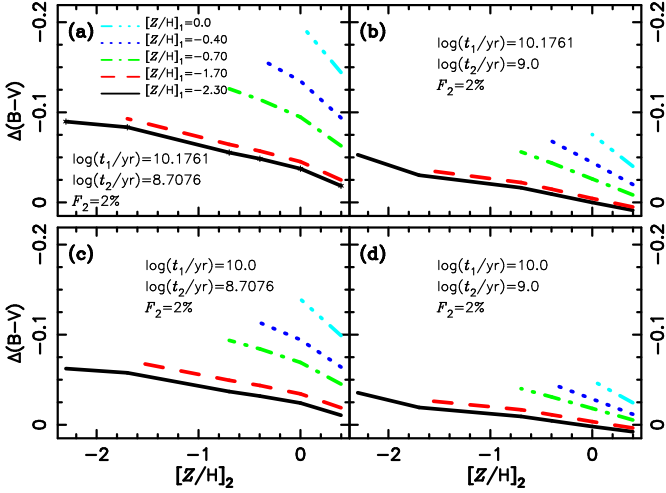}
      \caption{Difference between $(B-V)$ of a CSP and that of its
      DSP versus $Z_{\rm 2}$. Solid, dashed, dot-dashed, dotted,
      and dash-dot-dot-dot are for metallicities of DSPs ([$Z$/H]$_{\rm 1}$) of
      -2.30, -1.70, -0.70, -0.40, and 0.0, respectively.
              }
         \label{FigVibStab}
   \end{figure}
%

\section{Possible distributions of the differences between colors}

    The difficulty of getting accurate stellar ages and metallicities for the DSPs of galaxies via colors is that we get
    the colors of CSPs rather than those of DSPs.
    In this section, we analyze the possible distributions of
    differences between colors
    of CSPs and those of the DSPs mainly via a Monte Carlo technique.

\subsection{The composite stellar population sample}
    We construct a set of CSPs for the study via the
    following steps. First, we generate the ages and metallicities ($t_{\rm 1}$ and [$Z/{\rm H}$]$_{\rm 1}$) of
    DSPs in our sample CSPs according to two distributions derived from Gallazzi et al.
    (\cite{gallazzi05}), who determined stellar ages and metallicities of 175\, 128 galaxies drawn from the Sloan Digital Sky Survey Data Release Two
    (SDSS-DR2) by comparing D4000, H$\beta$, H$\delta_A$+H$\gamma_A$, [Mg$_{\rm 2}$Fe], and [MgFe]$'$ indices of galaxies
    to predictions of SSPs of the BC03.
    In particular, all galaxies have a ratio
    of signal-to-noise S/N $\ge$ 20 and do not have current star
    formation. Stellar ages used in this work are corrected for the light travel time
    ($H_{\rm 0}$ = 70 km s$^{\rm -1}$ {Mpc}$^{\rm -1}$). This will not obviously change our result, because the colors of CSPs
    are more sensitive to the age of the YSP, $t_{\rm 2}$, rather than that of the
    DSP, $t_{\rm 1}$. However, the correction can increase the difference between
    $t_{\rm 1}$ and $t_{\rm 2}$ slightly, and, correspondingly, the difference between colors of a CSP and its DSP.
    This makes the difference calculated in this work closer
    to the actual difference between colors of galaxies and their DSPs,
    as galaxies possibly contain more than two SSPs.
    Second, the metallicity of the YSP of a CSP is taken as the same as that of the DSP.
    Third, the ages of YSPs are given randomly by a Monte Carlo
    process, with a natural assumption that the YSP of a CSP is younger than the DSP of the CSP (0 $<$ $t_{\rm 2}$ $<$ $t_{\rm 1}$).
    Fourth, the mass fraction ($F_{\rm 2}$) of YSP is calculated according to the ages of
    the DSP and YSP by the equation

    \begin{equation}
     \it
     F_{\rm 2} = F_{\rm 0}~.~{\rm exp}~[\frac{t_{\rm 2}-t_{\rm
     1}}{\tau}] ,
    \end{equation}
    which assumes that the mass fraction of the young population
    declines exponentially with decreasing age of the YSP, according to some studies about
    the star formation histories of galaxies, e.g., Bruzual
    (\cite{bruzual83}) and Thomas et al. (\cite{thomas05}).
    The variables $t_{\rm 2}$ and $t_{\rm
    1}$ are the ages of the YSP and DSP of a CSP, respectively. The value of $F_{\rm 0}$ is the
    maximum mass fraction, while $\tau$ is a free parameter. For a standard case, we take 0.5 and 3.02 for $F_{\rm 0}$ and
    $\tau$, respectively, which gives about 15\% CSPs with $F_{\rm 2}$ $\leq$
    4\%. This agrees with the result of Yi et al. (\cite {yi05}) that
    there are about 15\% bright early type galaxies with recent ($\leq$ 1 Gyr) star formation at a level more than 1\% -- 2\% in mass
    compared to the total stellar mass.
    In this way, 30\, 000 CSPs with $t_{\rm  1}$ $\geq$ 1 Gyr are constructed. The distributions
    of the input parameters of these CSPs are shown in Fig. 8.

   \begin{figure}
   \centering
   \includegraphics[angle=-90,width=88mm]{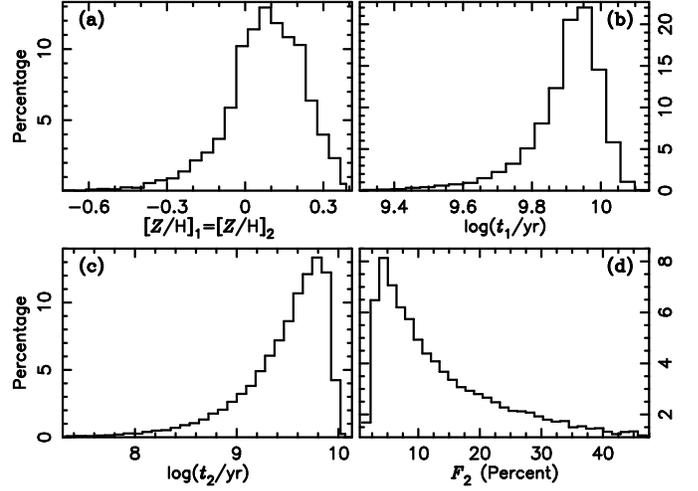}
      \caption{Distributions of five input parameters of our sample
      CSPs. Panels (a), (b), (c), and (d) show the distributions of
      [$Z$/H]$_{\rm 1}$ and [$Z$/H]$_{\rm 2}$, log($t_{1}\rm /yr$), log($t_{2}\rm /yr$), and $F_{2}$, respectively.
              }
         \label{FigVibStab}
   \end{figure}
%


\subsection{Distributions of differences between colors of CSPs and those of their DSPs}
    We investigate the distributions of the differences between colors of CSPs and those of their DSPs using the above sample. The results of
    eight colors, which are shown to be relatively sensitive to stellar age or metallicity (see Li et al. \cite{li07}),
    are plotted in Figs. 9 and 10. The differences are calculated by
    Eq. (1).
    We see that more CSPs exhibit bluer colors than their DSPs.
   \begin{figure}
   \centering
   \includegraphics[angle=-90,width=88mm]{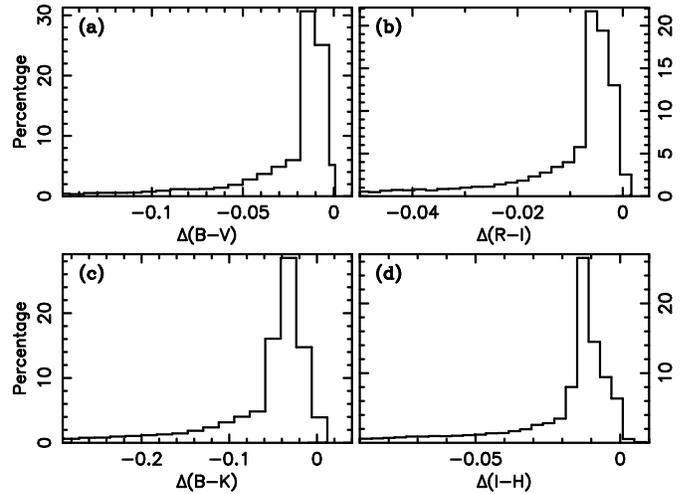}
      \caption{Distributions of differences between colors of CSPs and those of their
      DSPs. Panels (a), (b), (c), and (d) show the distributions for
      $(B-V)$, $(R-I)$, $(B-K)$, and $(I-H)$, respectively. $\Delta (B-V)$, $\Delta (R-I)$,
      $\Delta (B-K)$, and $\Delta (I-H)$ are calculated by Eq.
      (1).
              }
         \label{FigVibStab}
   \end{figure}
%
   \begin{figure}
   \centering
   \includegraphics[angle=-90,width=88mm]{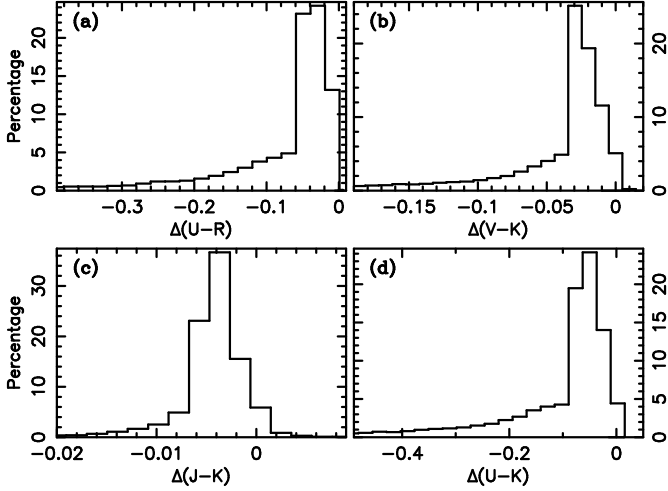}
      \caption{Similar to Fig. 9, but for $(U-R)$ (Panel a), $(V-K)$ (Panel b), $(J-K)$ (Panel c), and $(U-K)$ (Panel d).
              }
         \label{FigVibStab}
   \end{figure}
%

\section{Possible corrections for the deviations in stellar ages and metallicities determined by colors}

\subsection{Distributions of these deviations}
    We investigate how the stellar ages and
    metallicities measured by colors deviate from those of the DSPs in the sample generated in Sect. 4.
    Because the two stellar-population parameters of galaxies are usually estimated via two
    colors due to the limitation of the observational data, we study the results derived from different color pairs in the work.
    In detail, stellar ages and metallicities are determined by [$(B-V)$,
    $(V-K)$], [$(B-V)$, $(B-K)$], and [$(B-V)$, $(I-H)$], respectively, using the BC03 SSPs.
    The three pairs of colors (hereafter color pairs) are selected by finding
    the best pairs for breaking the stellar age--metallicity degeneracy (Worthey \cite{worthey94}) ``by eye'' via color--color grids of color pairs constructed
    via three metallicity-sensitive colors, $(B-K)$,
    $(I-H)$, $(V-K)$, and three age-sensitive colors, $(B-V)$, $(U-R)$, and $(R-I)$ (see Li et al.
    \cite{li07}). Note that [$(B-V)$, $(V-K)$] and [$(B-V)$, $(B-K)$] are equivalent from a mathematical view, as they
    relate to the same three magnitudes, but they have different abilities for
    disentangling the well-known degeneracy. For example,
    it seems impossible to disentangle the above degeneracy via [$(B-K)$, $(V-K)$].
    The reason is that both $(B-K)$ and $(V-K)$ are relatively sensitive to metallicity, but
    their relative metallicity sensitivities are different (Li et al. \cite{li07}).
    Thus it is necessary to compare the results derived from
    [$(B-V)$, $(B-K)$] and [$(B-V)$, $(V-K)$].
    Then the deviations  ($\Delta$${\rm Age}$ and $\Delta$${\rm Z}$) of the best-fitted values (age,
    $t_{\rm f}$ and metallicity, $Z_{\rm f}$)
    from those of their DSPs (age, $t_{\rm 1}$ and metallicity, $Z_{\rm 1}$) are calculated
    ($\Delta$${\rm Age}$ = $t_{\rm f}$ - $t_{\rm 1}$,
    $\Delta$${\rm Z}$ = $Z_{\rm f}$ - $Z_{\rm 1}$). The
    distributions of the deviations in stellar ages and metallicities are plotted in Fig. 11, in which the
    results reported by three different pairs of colors are shown, respectively. As we see, the results
    derived from fitting three different pairs of colors are similar.
    They have similar shapes and all of them peak near $\Delta$Age = -0.8 Gyr and $\Delta$Z =
    0.0002, with averages -2.14 Gyr and 0.0027 in stellar age and metallicity, respectively.
    For the convenience of using these distributions in future works, we fit them by equations:
\begin{equation}
    \it {\rm P} ({\rm \Delta} {\rm Age})=P_{\rm 0} + \sum_{\rm i=1}^{\rm 3}{A_{\rm i}~{\rm exp~}
    [\frac{({\rm \Delta} {\rm t} - X_{\rm i})^{\rm 2}}{{\rm -2}W_{\rm i}^{\rm 2}}
    ]},
\end{equation}

\begin{equation}
    \it {\rm P} ({\rm \Delta} {\rm Z})=P_{\rm 0}{'} + \sum_{\rm i=1}^{\rm 3}{A'_{\rm i}~{\rm exp~}
    [\frac{({\rm \Delta} {\rm Z} - X{'}_{\rm i})^{\rm 2}}{{\rm -2}W{'}_{\rm i}^{\rm 2}} ]},
\end{equation}
    and the coefficients are listed in Table 2. Here P($\Delta$Age) and
    P($\Delta$Z) in the two equations are the percentages probabilities. The two equations are normalized, respectively, by
\begin{equation}
    \rm \int P (\Delta Age)~d(\Delta Age) = 1,
\end{equation}
 and
\begin{equation}
    \rm \int P (\Delta Z)~d(\Delta Z) = 1.
\end{equation}
    To see the fits of the two distributions
    clearly, we plot the distributions derived from fitting $(B-V)$ and
    $(B-K)$ colors
    and the distributions calculated by Eqs. (3) and (4) in Fig. 12.
    In practice, these distributions can possibly be used to correct for the
    deviations of
    stellar ages and metallicities determined by colors and get
    some reliable estimations for the two parameters.

\begin{table}[]
\caption[]{Coefficients of Eqs. (3) and (4).} \label{Tab:1}
\begin{center}\begin{tabular}{cccc}
\hline\noalign{\smallskip}
\it P$_0$ &\it A$_1$ &\it X$_1$  &\it W$_1$\\
\noalign{\smallskip}
0.0029 &11.9650 &-0.7565 &0.6375\\
\noalign{\smallskip}
&\it A$_2$ &\it X$_2$ &\it W$_2$ \\
&2.3850 &-2.7804  &0.8500\\
\noalign{\smallskip}
&\it A$_3$ &\it X$_3$ &\it W$_3$\\
\noalign{\smallskip}
&1.1427 &-5.3102 &1.5938\\
\hline\hline\noalign{\smallskip}
\it P$_0$$~'$ &\it A$~'$$_1$ &\it X$~'$$_1$ &\it W$~'$$_1$ \\
\noalign{\smallskip}
0.0016 &33.6151 &0.0004 &0.0007\\
\noalign{\smallskip}
&\it A$~'$$_2$ &\it X$~'$$_2$ &\it W$~'$$_2$ \\
\noalign{\smallskip}
&4.1408 &0.0025  &0.0019\\
&\it A$~'$$_3$ &\it X$~'$$_3$  &\it W$~'$$_3$ \\
&0.5217 &0.0168 &0.0069\\

\noalign{\smallskip}\hline
\end{tabular}\end{center}
\end{table}

   \begin{figure}
   \centering
   \includegraphics[angle=-90,width=88mm]{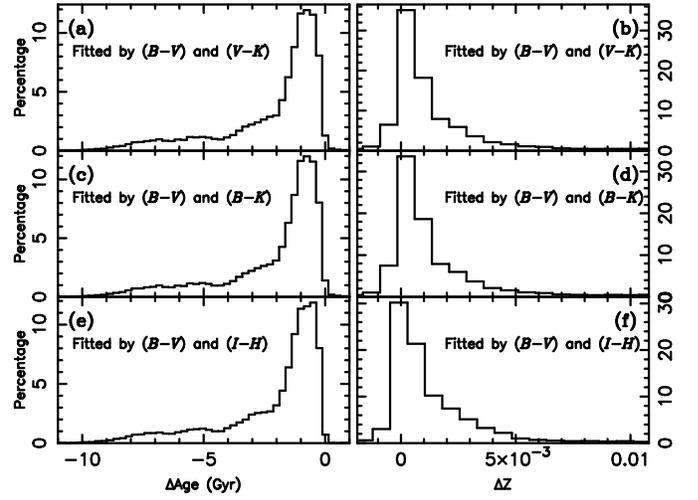}
      \caption{Distributions of the differences between stellar parameters determined
      by fitting color pairs and those of the DSPs of CSPs ($\Delta$${\rm Age}$ = $t_{\rm f}$ - $t_{\rm 1}$,
    $\Delta$${\rm Z}$ = $Z_{\rm f}$ - $Z_{\rm 1}$).
              }
         \label{FigVibStab}
   \end{figure}
%
   \begin{figure}
   \centering
   \includegraphics[angle=-90,width=88mm]{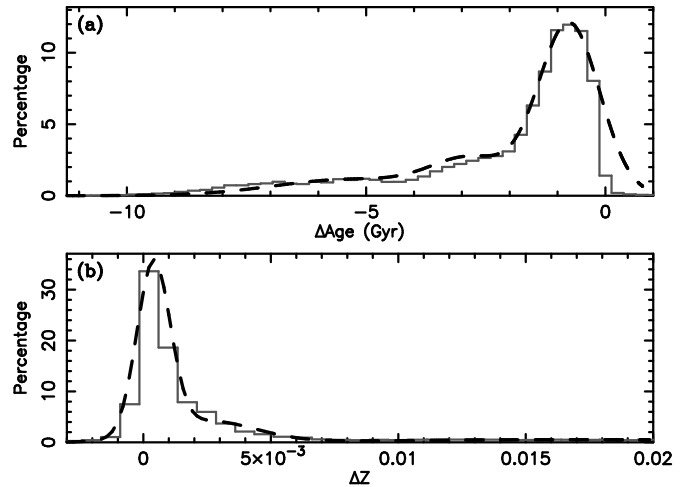}
      \caption{The fitted distributions of $\Delta \rm {Age}$ and $\Delta Z$.
      $\Delta \rm {Age}$ and $\Delta Z$ have the same meanings as in Fig. 11.
      Solid lines show the distributions derived from a
      sample of CSPs and dashed lines show the distributions calculated by
      Eqs.
      (3) and (4).
      Panels (a) and (b) show the distributions for,
      respectively,
      the differences in stellar age and in metallicity determined by [$(B-V)$
      , $(B-K)$].
              }
         \label{FigVibStab}
   \end{figure}
%

\subsection{Correcting the deviations of stellar ages and metallicities}
    We tried to correct for the deviations in stellar ages and
    metallicities determined by fitting a pair of
    colors for a sample of 2\, 000 CSPs with $t_{\rm 1}$ $>$ 3 Gyr.
    The sample is generated by the same method as in Sect. 4 and all of their $UBVRIJHK$ colors are calculated.
    We get the best-fitted stellar ages and metallicities ($t_{\rm f}$ and $Z_{\rm f}$) for CSPs by fitting their $(B-V)$ and $(B-K)$ colors. The distributions of the differences between the best-fitted parameters and those of
     their DSPs ($\Delta {\rm Age}$ = $t_{\rm f}$ - $t_{\rm 1}$ and
    $\Delta {\rm (Fe/H)}$ = [$Z/{\rm H}$]$_{\rm f}$ - [$Z/{\rm H}$]$_{\rm 1}$) are then calculated, and are shown in panels (a) and (b) of Fig. 13.
    We see that some stellar ages and metallicities younger and richer than those of the DSPs of CSPs are
    determined by color pairs. This suggests that without any corrections, it would be difficult to get further reliable
    results based on the stellar ages and metallicities determined by color
    pairs.

    To correct for the deviations in stellar ages
    and metallicities derived from colors, we tried two methods.
    In the first, we corrected the stellar ages and metallicities derived from colors using two
    maximum-likelihood values, i.e., -0.8 Gyr and 0.0002 for stellar age and metallicity, respectively.
    The results show that about 50\% CSPs
    have stellar ages and metallicities close to those of the DSPs of these populations after the correction
    ($\Delta {\rm Age}$ $<$ 1 Gyr, $\Delta {\rm ([Fe/H])}$ $<$ 0.05).
    In the second, we corrected stellar ages with the distribution of $\Delta$Age,
    which is shown in Fig. 12. Each corrected age was obtained by subtracting a random value that fits to the
    distribution shown in panel (a) of Fig.12 from the best-fitted age, $t_{\rm f}$. Stellar metallicities
    are corrected in a similar way, using the distribution shown in panel (b) of Fig.12.
    We call the correction ``distribution
    correction''. For  convenience, we use the distributions obtained by
    this work rather than ones calculated by Eqs. (3) and (4).
    The distribution of the differences between corrected
    stellar ages and those of their corresponding DSPs can be seen in the left column of Fig.
    18, and Fig. 12 is useful for comparison.
    We see clearly that the stellar ages corrected by us distribute homogeneously around those of their DSPs.
    A similar correction is also used to correct the
    stellar metallicities and the results shown in the right column of Fig.
    18. As can be seen, the metallicities of CSPs are homogeneously
    distributed
    around those of their DSPs after the correction (see Fig. 12 for comparison).
    Therefore, the distribution correction
    can help us to get stellar ages and metallicities distributed around those of the DSPs of CSPs.
    To see the effects of the correction more clearly, we plot the results in another way (Fig. 13). In the plot, uncorrected stellar ages, log($t_{\rm f}/{\rm
    yr}$), uncorrected stellar metallicities, [$Z/{\rm H}$]$_{\rm f}$=log($Z_{\rm f}/Z_\odot$), corrected stellar ages, log($t_{\rm c}/{\rm
    yr}$), and corrected stellar metallicities, [$Z/{\rm H}$]$_{\rm c}$=log($Z_{\rm c}/Z_\odot$) of 2\, 000 CSPs are plotted versus those of their DSPs (log($t_{\rm 1}/{\rm yr}$) and [$Z/{\rm H}$]$_{\rm 1}$).
    Most stellar
    ages and metallicities corrected by us fit to those of their DSPs when taking
    possible uncertainties in stellar age (1.5 Gyr) and metallicity (0.005) into account.
    Therefore, the averages of stellar ages
    and metallicities of a sample of galaxies can be estimated in this way, as can their distributions of stellar ages and metallicities.

    We apply the correction to the results
    of 53 elliptical galaxies, whose stellar ages and metallicities are estimated by $(B-V)$ and $(B-K)$
    colors by Li et al. (\cite{li07}).
    The distributions of uncorrected and corrected stellar parameters are plotted in Fig. 14.
    After the correction, stellar populations of these galaxies, which average about
    6 Gyr and 0.0375 in stellar age
    and metallicity, respectively, come to be older
    and less metal rich than the uncorrected ones. The results seem to be
    closer to the typical values of early type galaxies that are determined by Lick/IDS indices or spectra (see, e.g.,
    Gallazzi et al. \cite{gallazzi05}), although these populations are somewhat younger and more metal
    rich.

    To check the reliability of stellar ages and metallicities
    measured by colors, we fit the two parameters for 61 early type
    galaxies respectively by two line-strength indices (H$\beta$ and [MgFe]=$\sqrt{\rm Mgb \times(Fe5270+Fe5335)/2}$~, see Gonz$\acute {a}$lez \cite{gonzalez93}) and
    two colors ($B-V$ and $B-K$), and compare the two kinds of
    results. Although there is a new definition of [MgFe], in which Fe5270 and Fe5335 are weighted as 0.72 and 0.28,
    respectively (Thomas et al. \cite{thomas03}), we take
    the the above definition, because we only have the data for $<{\rm Fe}>$ = (Fe5270+Fe5335)/2.
    The Lick indices of the sample galaxies are selected from Sil'chenko (\cite{sil'chenko06}), Trager
    (\cite{trager00}), and Thomas et al. (\cite{thomas05}), which can also be found from
    Kuntschner et al. (\cite{kuntschner06}). The photometry data and redshifts of these galaxies are
    obtained from the NASA/IPAC Extragalactic Database (NED) and HyperLeda database (http://leda.univ-lyon1.fr/), respectively. The $K$-band
    magnitudes are calculated from $K_{\rm s}$-band data and
    take the $k$-correction and galactic extinction into account.
    Here, we take the methods presented by Bessell
    (\cite{bessell05}) \&
    Girardi et al. (\cite{girardi03}) for transforming $K_{\rm s}$-band data into $K$-band
    and $k$-correcting the data. The galactic extinction is
    calculated using the calculator supplied by
    NED.

    We list the stellar ages and metallicities fitted by the two
    methods in Table 3. Note that we only show 18 galaxies that have best-fitted stellar
    ages ($t_{\rm lick}$ and $t_{\rm color}$) smaller than 15 Gyr here,
    according to the age of the universe (see, e.g., Shafieloo et al. \cite{shafieloo06}).
    The results in the table were not corrected for the effects of YSPs, so that we can easily estimate
    the difference between the stellar-population parameters determined by colors and those by line-strength indices.
    We did not correct for the light travel time either,
    because the correction is not important for nearby galaxies.
    In Fig. 15, we compare the
    stellar parameters determined by colors and corrected via ``distribution correction'' with those determined by
    line strength indices. Note that the stellar ages and metallicities derived from
    colors are corrected in the whole sample of 61 galaxies.
    Possible uncertainties (1.5 Gyr in age and 0.005 in metallicity) are shown in Fig. 15.
    We see that the two kinds of results are consistent,
    taking uncertainties about 2.5 Gyr in age and 0.01 in
    metallicity.

    When we try to compare the stellar ages and metallicities estimated respectively
    by
    a color pair, [$(B-V)$, $(B-K)$], and four independent colors, $(B-V)$, $(B-K)$, $(I-H)$,
    and $(B-I)$, we obtain similar results. The colors of galaxies are selected from Michard (\cite{michard05}). The best-fitted stellar ages and metallicities
    are found by minimizing the function
\begin{equation}
    \it Q~(j)= \sqrt{\frac{\sum_{i=1}^{n}{(C_{i,j}^{'}-C_{i})^{\rm 2}}}{n}},
\end{equation}
    where $C_{i}$ is the $i$th observed color, $C_{i,j}^{'}$ the color corresponding to $C_{i}$ for the
    $j$th SSP, and $n$ (2 or 4) the number of colors used for fitting.
    In the fitting, we did not take the observational
    uncertainties into account, because there is no reliable data.
    The possible ranges of stellar ages and metallicities determined by a color
    pair and by four independent colors are given at 1 $\sigma$ (68.3\%) and 2 $\sigma$ (95.4\%) confidence levels.
    The confidence levels can be estimated via a chi-square test, taking a typical uncertainty (0.02 mag) for each color.
    The detailed comparison are seen in Fig. 16,
    in which the results of four elliptical galaxies are shown.
    From the figure, we see that the possible
    stellar ages and metallicities are grouped along the line of the well-known age--metallicity degeneracy (Worthey \cite{worthey94}),
    and the possible ranges determined by the two
    methods are similar. This can be interpreted as there being no color
    sensitive only to DSP or YSP among the four colors. One can refer to Li et al. (\cite{li07}),
    in which a principal component analysis (PCA) of colors for SSPs with various ages was
    shown. However, most of the best-fitted stellar ages and metallicities determined by four colors are
    respectively younger and more metal rich than those determined by a color pair, as we see. This suggests that
    the YSPs in galaxies affect the determination of DSPs of galaxies more strongly when we measure
    stellar ages and metallicities by fitting a few independent colors.

   \begin{figure}
   \centering
   \includegraphics[angle=-90,width=88mm]{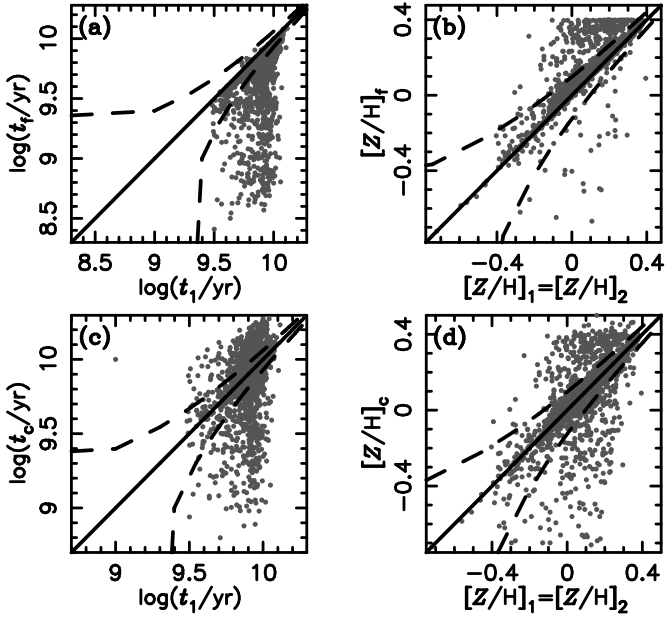}
      \caption{Plots of uncorrected and corrected stellar ages and
      metallicities versus those of their corresponding DSPs. Panels (a)
      and (b) show the plot of the uncorrected stellar
      parameters (log($t_{\rm f}/{\rm yr}$) and [$Z/{\rm H}$]$_{\rm f}$) and those of their corresponding DSPs (log($t_{\rm 1}/{\rm yr}$) and [$Z/{\rm H}$]$_{\rm 1}$).
      The uncorrected stellar ages and metallicities are determined by $(B-V)$ and
      $(B-K)$ colors. Panels (c)
      and (d) show the similar relations for distribution-corrected
      stellar ages and metallicities (log($t_{\rm c}/{\rm yr}$) and [$Z/{\rm H}$]$_{\rm c}$). Dashed lines in panels (a) and (c) show a $\pm$ 1.5 Gyr
      spread about the unity (solid) line for stellar ages. In panels (b) and (d), they
      show and a $\pm$ 0.005 spread
      about the unity (solid) line for stellar metallicities.
              }
         \label{FigVibStab}
   \end{figure}
%

   \begin{figure}
   \centering
   \includegraphics[angle=-90,width=88mm]{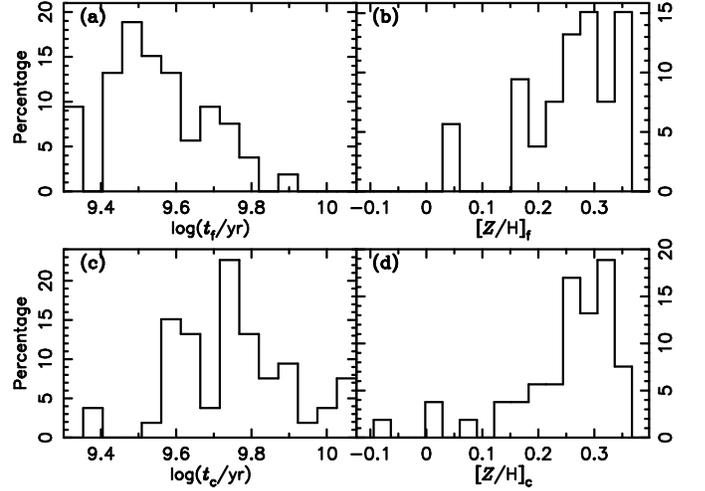}
      \caption{Distributions of uncorrected and corrected stellar
      ages and metallicities of 53 elliptical galaxies. Panels (a)
      and (b) show the distributions of the uncorrected results (log($t_{\rm f}/{\rm yr}$) and [$Z/{\rm H}$]$_{\rm f}$) while panels (c) and (d)
      show those for the corrected results (log($t_{\rm c}/{\rm yr}$) and [$Z/{\rm H}$]$_{\rm c}$).
              }
         \label{FigVibStab}
   \end{figure}
%

\begin{table}[]
\caption[]{Stellar ages and metallicities fitted respectively by
line strength indices (H$\beta$ and [MgFe]) and colors ($B-V$ and
$B-K$).} \label{Tab:1}
\begin{center}\begin{tabular}{crrrr}
\hline\noalign{\smallskip}
   Galaxy  &log(\it t\rm$_{\rm lick}$/yr) &[$\it Z$/H]$_{\rm lick}$& log(\it t\rm$_{\rm color}$/yr) &[$\it Z$/H]$_{\rm color}$ \\
\hline\noalign{\smallskip}
   NGC0080 &10.0019        & 0.1072   &9.6750           & 0.2707    \\
   NGC2681 & 9.5119        &-0.4815   &9.3900           &-0.1707    \\
   NGC2950 & 9.5119        & 0.3627   &9.4950           & 0.1917    \\
   NGC3166 & 9.5119        &-0.0066   &9.7650           &-0.0809    \\
   NGC3377 & 9.5969        & 0.1553   &9.9450           &-0.4377    \\
   NGC3384 & 9.9469        &-0.0022   &9.7600           & 0.0881    \\
   NGC3941 &10.1269        &-0.1612   &9.7700           & 0.0374    \\
   NGC4026 & 9.9769        & 0.0000   &9.6600           & 0.1492    \\
   NGC4150 & 9.7769        &-0.5302   &9.8550           &-0.5768    \\
   NGC4179 & 9.9169        & 0.1847   &9.5500           & 0.2055    \\
   NGC4350 & 9.9769        & 0.1173   &9.6700           & 0.2455    \\
   NGC4570 & 9.9519        & 0.1945   &9.6600           & 0.2201    \\
   NGC5308 & 9.9369        & 0.2822   &9.5300           & 0.2240    \\
   NGC5574 & 9.5119        &-0.0757   &9.5500           &-0.1397    \\
   NGC7013 & 9.7319        &-0.1457   &9.4300           &-0.6289    \\
   NGC7280 & 9.9269        &-0.2147   &9.4700           & 0.1055    \\
   NGC7457 & 9.6519        &-0.1457   &9.9350           &-0.4685    \\
   NGC7743 & 9.6869        &-0.1367   &9.4750           & 0.0128    \\


\noalign{\smallskip}\hline
\end{tabular}\end{center}
\begin{list}{}{}
\item[$Note:$] Suffixes ``lick'' and ``color'' represent the results fitted
by line strength indices and those by colors, respectively.
\end{list}
\end{table}

   \begin{figure}
   \centering
   \includegraphics[angle=-90,width=88mm]{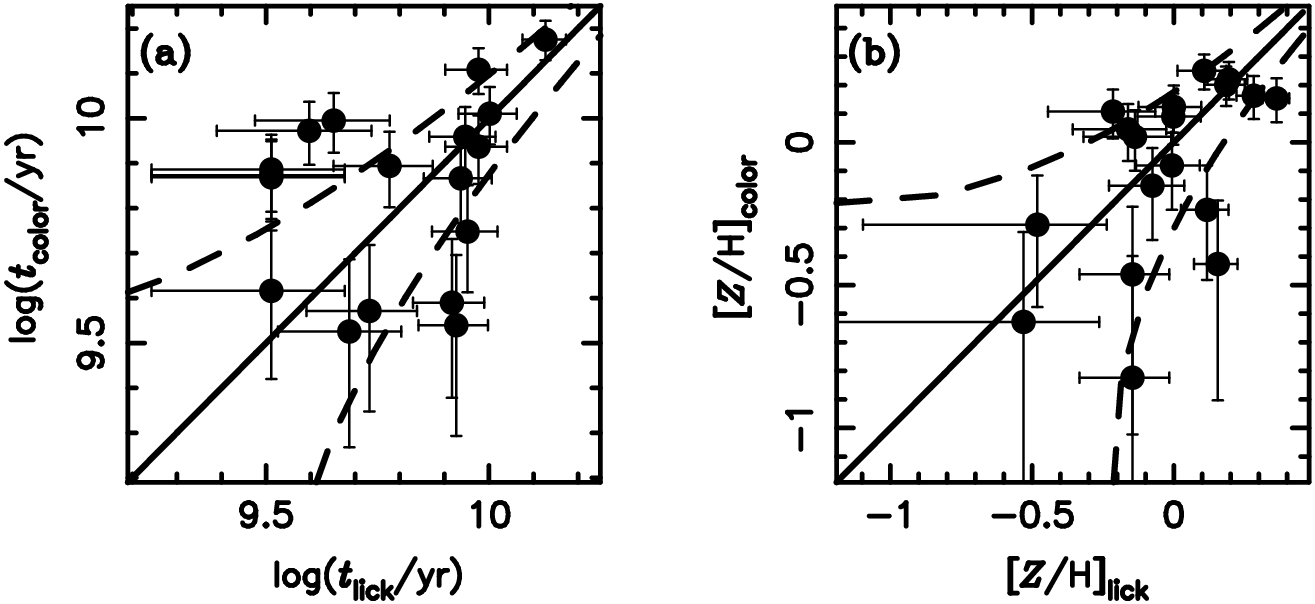}
      \caption{Stellar ages and
      metallicities fitted by [$(B-V),(B-K)$] versus those fitted by H$\beta$ and [MgFe].
      Dashed lines show a $\pm$ 2.5 Gyr
      spread about the unity (solid) line for stellar ages in panel (a), while they show a $\pm$
      0.01 spread in stellar metallicities in panel (b). Error
      bars show uncertainties of 1.5 Gyr and 0.005 in stellar age
      and metallicity, respectively.
              }
         \label{FigVibStab}
   \end{figure}
%

\subsection{Effect of the calculation of mass fractions of young populations}
    We took 0.5 and 3.02 for $F_{\rm 0}$ and
    $\tau$, respectively, to calculate the mass fractions of YSPs of CSPs via Eq. (2) in our standard case,
    but different values may lead to different distributions of deviations in stellar ages and metallicities.
    Here we test how $F_{\rm 0}$ and $\tau$ change the distributions of deviations in the stelllar-population parameters estimated by two
    colors. Given $\tau$ = 3.02, we test the effects of $F_{\rm 0}$
    via four values, i.e., 0.2, 0.3, 0.4, and
    0.5. The
    distributions relating to different values of $F_{\rm 0}$ are
    plotted in panels (a) and (b) of Fig. 17. Similarly, given $F_{\rm 0}$ = 0.5, we test the effects of $\tau$, taking 2.0, 3.02, 5.0, and 10.0.
    The results are shown
    in panels (c) and (d) of Fig. 17. In addition, we test the distributions
    by taking fixed values (5\%, 10\%, and 15\%) for $F_{\rm 2}$, which are shown in panels (e) and (f) of Fig. 17.
    As we see, high values of $F_{\rm 0}$
    or $\tau$ will lead to more populations with big deviations in two stellar-population parameters. However, the distributions of deviations in the two stellar-population
    parameters are similar for different formulations of $F_{\rm 2}$.
    We tried to correct for the deviations using the distributions
    obtained in the standard case, by a few samples including 2\, 000 CSPs
    generated by different formulations of $F_{\rm 2}$.
    The distributions of differences between
    two corrected stellar-population parameters and those of the DSPs are shown in Fig. 18. In the
    test, samples of CSPs are generated by taking different formulations.
    As we see, the stellar ages and metallicities distribute around
    those of DSPs, after the distribution correction. Therefore,
    even if the star formation histories are different,
    the averages of the ages and metallicities of the DSPs of galaxies can
    be estimated from their colors, by using the the distributions obtained in the standard case to correct the
    results, as can the distributions of the two stellar-population parameters.


\begin{figure*}
\centering
\includegraphics[angle=-90,width=12cm]{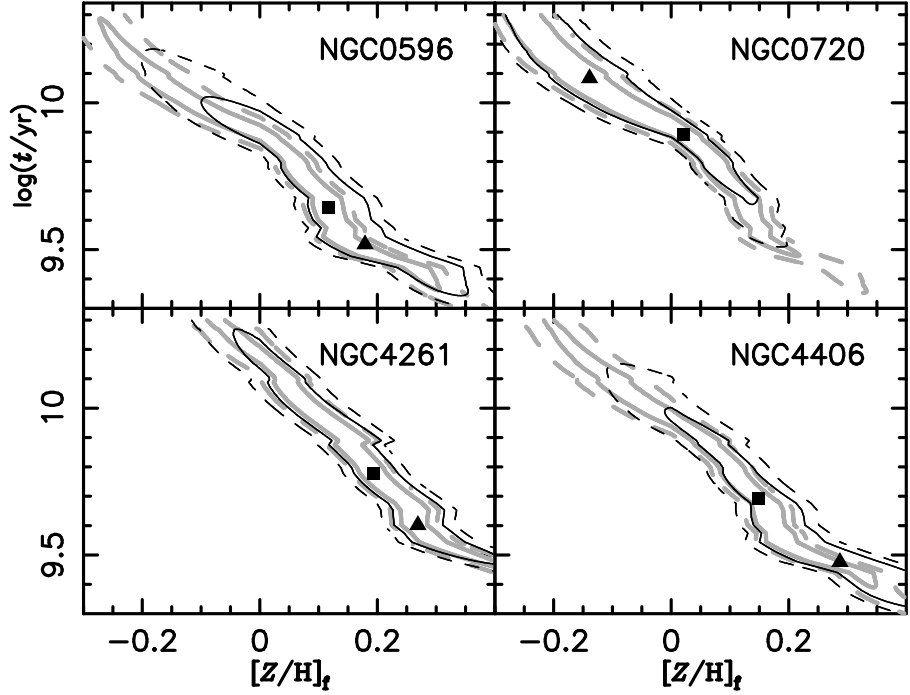}
\caption{Comparison of the stellar parameters (age and metallicity)
determined by a color pair ($B-V$ and $B-K$) with those determined
by four independent colors ($B-V$, $B-K$, $B-I$, $I-H$). Squares and
triangles show the best-fit stellar ages and metallicities that are
determined by a color pair and four independent colors,
respectively. Gray and dark lines show the 1 $\sigma$ (solid) and 2
$\sigma$ (dashed) ranges of stellar ages and metallicities for the
two-color method and multi-color method, respectively.} \label{<Your
label>}
\end{figure*}


\begin{figure*}
\centering
\includegraphics[angle=-90,width=12cm]{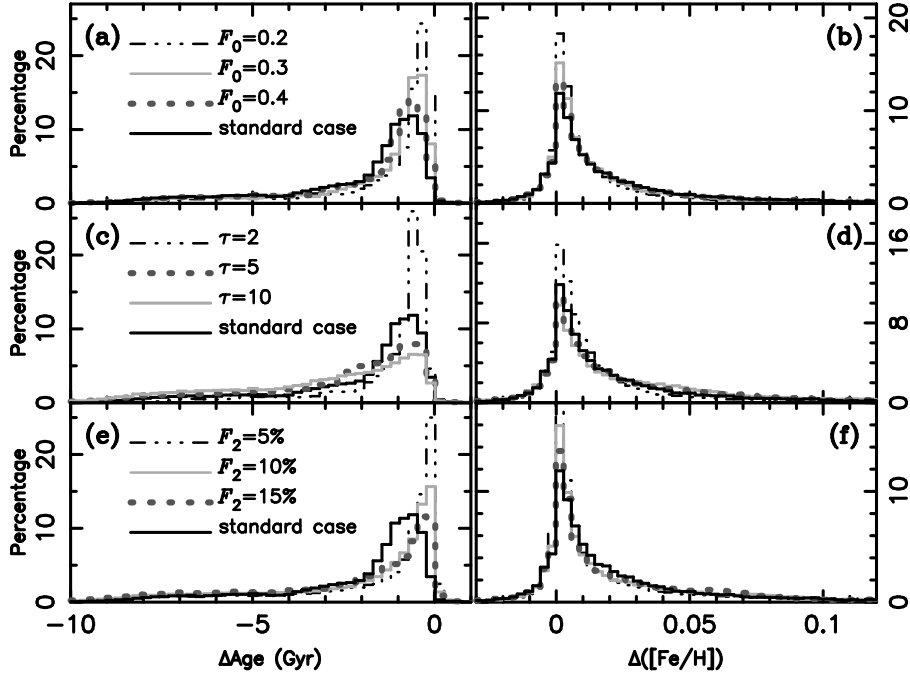}
\caption{Distributions of the deviations in stellar ages
      and metallicities determined by color pairs for
      different calculations of the mass fractions of YSPs of composite star
      systems. Panels (a, b), (c, d), and (e, f) show for varying $F_{\rm 0}$, $\tau$, $F_{\rm 2}$, respectively.
      In panels (a), (b), (c), and (d), $F_{\rm 2}$ are calculated by Eq. (2).
      Our standard case takes $F_{\rm 0}$=0.5 and $\tau$=3.02 for Eq. (2).} \label{<Your label>}
\end{figure*}

\begin{figure*}
\centering
\includegraphics[angle=-90,width=12cm]{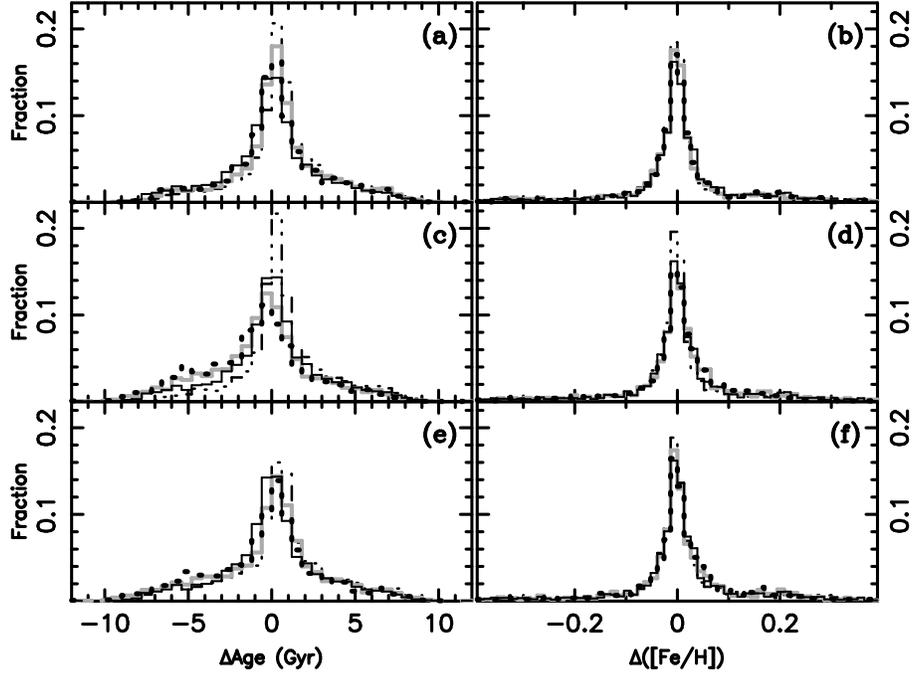}
\caption{Distributions of the deviations in corrected stellar ages
      and metallicities from those of their corresponding DSPs.
      Solid lines show for 2\, 000 CSPs with $F_{\rm 2}$
      calculated by Eq. (2) with $F_{\rm 0}$ = 0.5 and $\tau$ = 3.02.
      The other samples of CSPs in Panels (a) and (b) are generated by
      taking $F_{\rm 0}$ = 0.2 (dash-dot-dot-dot), 0.3 (gray solid), 0.4
      (dotted),
      and $\tau$ = 3.02 for Eq. (2).
      The ones in Panels (c) and (d) are generated by
      taking $\tau$ = 2 (dash-dot-dot-dot), 5 (gray solid), 10
      (dotted) and $F_{\rm 0}$ = 0.5 for Eq. (2).
      Those in Panels (e) and (f) are generated by
      taking $F_{\rm 2}$ = 5\% (dash-dot-dot-dot), 10\% (gray solid), 15\% (dotted),
      respectively.} \label{<Your label>}
\end{figure*}


\section{Conclusions and discussions}

    We first investigated the effects on colors of a
    composite stellar population (CSP) of three parameters (age, $t_{\rm 2}$, mass fraction, $F_{\rm 2}$,
    and metallicity, [$Z/{\rm H}$]$_{\rm 2}$) of the young stellar population (YSP). Then we
    studied the potential distributions of the deviations of stellar ages and
    metallicities determined by color pairs, based on a set
    of CSPs that are rebuilt via SSPs of the BC03.

    Our results show that the age ($t_{\rm 2}$) and mass
    fraction ($F_{\rm 2}$) of the YSP affect colors of the system significantly, while the
metallicity of YSP ($Z_{\rm 2}$ or [$Z/{\rm H}$]$_{\rm 2}$)
    affect the system's colors more weakly. Thus if there were YSPs in galaxies and they were not too old (e.g., $t_{\rm 2}$ $<$ 5
    Gyr), the
    colors of galaxies should be significantly different from those of their DSPs. This means that it is difficult to
    get the accurate ages and metallicities of the DSPs of galaxies by comparing colors of galaxies to those
    of SSPs directly. In addition, the effects of the
    age ($t_{\rm 2}$) and mass fraction ($F_{\rm 2}$) of the YSP are found to be degenerate:
    The smaller the age or greater mass fraction, the bigger the deviation in the colors of a CSP compared to
    those of the DSP of the system.
    But, as a whole, the effects of $t_{\rm 2}$ on the colors of star systems are stronger than those of $F_{\rm 2}$.
    Our results also show that the effects of the metallicity of the
    YSP ($Z_{\rm 2}$ or [$Z/{\rm H}$]$_{\rm 2}$) on the colors of a system are much weaker
    than those of $t_{\rm 2}$ and $F_{\rm 2}$.

    Furthermore, we give possible distributions for the deviations in the stellar ages and metallicities
    determined by color pairs in this work, which can help us to get reliable estimations of the averages and distributions of
    the stellar ages and metallicities of
    a sample of galaxies from their colors. This is possibly the most important result of this work. According
    to our results, the stellar ages and metallicities measured by color pairs, e.g.,
    [$(B-V)$, $(V-K)$], [$(B-V)$, $(B-K)$], and [$(B-V)$, $(I-H)$], are about 2.14 Gyr younger, while
    0.0027 more metal rich than those of the DSPs of CSPs on average. The differences between
    the colors of CSPs and those of their DSPs were also studied in this
    work.

    When we tried to study the stellar ages and metallicities determined respectively by a color
    pair and four independent colors, we got similar ranges for the two stellar-population parameters.
    The results also suggest that YSPs affect the determination of the ages and metallicities of DSPs more strongly
    when we estimate the two stellar-population parameters by a few independent colors rather than a color pair.

    Our results can be used widely, although we took the BC03 models
    for the study. The reason is that various models usually predict similar
    colors for the same stellar population. For example, $(V-K)$ colors predicted
    by PEGASE models are only significantly different from those predicted by the
    BC03 models when the stellar age is between 0.06 and 0.32 Gyr,
    but $(B-V)$ colors predicted by the two models are very similar for all SSPs (Bruzual \& Charlot
    \cite{bruzual03}).

    In addition, the results can help us not only estimate the
    stellar ages and metallicities but also study the fundamental
    relations, such as the fundamental plane (FP) (Djorgovski \& Davis \cite{djorgovski87})
    and Kormendy relation (Kormendy \cite{kormendy77}) of elliptical galaxies and the bulges of lenticulars,
    which are powerful distance indicators for galaxies and important constraints on
    galaxy formation and evolution.
    According to our results, YSPs increase B luminosity, and, correspondingly, B
    surface brightness, compared to that of the DSPs of galaxies. In other words, YSPs affect
    the light of galaxies significantly. However, YSPs
    are unimportant for affecting the dynamic parameters of
    galaxies, as they only contain a small fraction of the stellar mass. Therefore, it is necessary to take
    the effects of the YSPs of galaxies into account when we study the fundamental relations
    relating to the dynamic parameters of galaxies, e.g., the relations transformed from
    the virial theorem. The FP of early type galaxies is just one of these kinds of relations.
    Using the stellar populations of 18 galaxies, which are
    measured from two colors ($B-V$ and $B-K$) and two line strength
    indices (H$\beta$ and [MgFe]), we calculated the difference
    between B surface brightness of a CSP and that of its DSP. The results
    show that the difference between B surface brightness of a CSP and its DSP can be as large as 1 mag,
    with an average of 0.33 mag. Therefore, taking only the DSPs in early type galaxies into account,
    the FP of early type galaxies will change:
    A pair of effective radius ( $r_{\rm e}$) and velocity dispersion ($\sigma$) will
    correspond to fainter B surface brightness compared to the FP obtained before. In this case, we
    will measure distance moduli smaller by about 0.33 mag for galaxies than those
    determined via the FP obtained before. Furthermore, if we correct for the effects of YSPs,
    the FP obtained in various bands may change. Possibly, the differences among FPs obtained in
    various bands may decrease, compared to the results of Bernardi et al. (\cite{bernardi03}).
    The Kormendy relation (Kormendy \cite{kormendy77}) will change and affect the
    determination of the distances of early type galaxies with a trend similar to the FP
    if we take the effects of YSPs into account.
    A given effective radius corresponds to fainter B surface brightness
    and then give us a smaller distance than the previous relation.
    Moreover, we find that the average of the difference between
    B surface brightness of a CSP and that of its DSP, i.e., 0.33 mag,
    is close to the typical scatter of the FP and Kormendy relation
    of elliptical galaxies and the bulges of lenticulars (0.3--0.5 mag)
    (see, e.g., Falc\'{o}n-Barroso et al. \cite{falcon02}; Djorgovski \& Davis \cite{djorgovski87}; Kormendy \cite{kormendy77}; Reda et al. \cite{reda05}).
    It implies that YSPs in early type galaxies may contribute to the scatter of the two fundamental relations.
    Thus the scatter in the two fundamental relations may be smaller if we take the effects of YSPs into account.

    In conclusion, the results of the work can help us to
    get stellar ages and metallicities close to those of the DSPs
    of galaxies via photometry data
    and to understand the fundamental relations of early type galaxies better.
    The main results of this work are available in electronic form
    at the CDS.

\begin{acknowledgements}
We gratefully acknowledge the anonymous referee for useful comments
and greatly improving the presentation of the paper, Dr.~ Anna
Gallazzi and her group for supplying us their results of some
galaxies, and Dr. Richard Simon Pokorny for checking the English. We
also thank the NASA/IPAC Extragalactic Database (NED) and HyperLeda
database (http://leda.univ-lyon1.fr/) for supplying us with the
photometry data of some galaxies. This work is supported by the
Chinese National Science Foundation (Grant Nos 10433030, 10521001),
and the Chinese Academy of Science (No. KJX2-SW-T06).
\end{acknowledgements}

\end{document}